\newcommand{\be}{\begin{equation}}
\newcommand{\ee}{\end{equation}}
\newcommand{\rstar}{\ensuremath{R_\star}}
\newcommand{\mstar}{\ensuremath{M_\star}}
\newcommand{\rpl}{\ensuremath{R_{p}}}
\newcommand{\mpl}{\ensuremath{M_{p}}}
\newcommand{\dotw}{\ensuremath{\dot{\omega}_{\rm GR}}\ } 
\documentclass[onecolumn,apj]{emulateapj}

\slugcomment{Accepted for publication in ApJ}
\lefthead{JORD\'AN \& BAKOS}
\righthead{GR PRECESSION IN EXOPLANETS}

\begin{document}
 
\title{
	Observability of the General Relativistic Precession of Periastra in
	Exoplanets
}

\author{
	Andr\'es Jord\'an\altaffilmark{1,2,3} and
	G\'asp\'ar \'A. Bakos\altaffilmark{1,4}
}

\altaffiltext{1}{
	Harvard-Smithsonian Center for Astrophysics, 
	60 Garden St., Cambridge, MA 02138;
	ajordan@cfa.harvard.edu, gbakos@cfa.harvard.edu.
}
\altaffiltext{2}{Clay Fellow.}
\altaffiltext{3}{Departamento de Astronom\'{\i}a y Astrof\'{\i}sica, 
Pontificia Universidad Cat\'olica de Chile, Casilla 306, Santiago 22,
Chile.}
\altaffiltext{4}{NSF Postdoctoral Fellow.}

\begin{abstract}
The general relativistic precession rate of periastra in close-in
exoplanets can be orders of magnitude larger than the magnitude of the
same effect for Mercury.  The realization that some of the close-in
exoplanets have significant eccentricities raises the possibility that
this precession might be detectable. We explore in this work the
observability of the periastra precession using radial velocity and
transit light curve observations. Our analysis is independent of the
source of precession, which can also have significant contributions
due to additional planets and tidal deformations.
We find that precession of the periastra of the magnitude expected
from general relativity can be detectable in timescales of $\lesssim
10$ years with current observational capabilities by measuring the
change in the primary transit duration or in the time difference
between primary and secondary transits. Radial velocity curves alone
would be able to detect this precession for super-massive, close-in
exoplanets orbiting inactive stars if they have $\sim 100$ datapoints
at each of two epochs separated by $\sim 20$ years.
We show that the contribution to the precession by tidal deformations
may dominate the total precession in cases where the relativistic
precession is detectable.
Studies of transit durations with {\it Kepler} might need to take into
account effects arising from the general relativistic and tidal induced
precession of periastra for systems containing close-in, eccentric
exoplanets. Such studies may be able to detect additional planets with
masses comparable to that of Earth by detecting secular variations in
the transit duration induced by the changing longitude of periastron.
\end{abstract}

\keywords{celestial mechanics --- planetary systems}

\section{Introduction}

Following the discovery of an extra-solar planet around the solar type
star 51 Pegasi \citep{Mayor1995a} there has been rapid progress in the
detection and characterization of extra-solar planetary systems.
The very early discoveries have shattered our view on planetary
systems, as certain systems exhibited short periods (51 Peg), high
eccentricities \citep[e.g.\ 70 Virginis b,][]{Marcy1996}, and massive
planetary companions \citep[e.g.\ Tau Boo b,][]{Butler1997a}.

Interestingly, systems with all these properties combined
(i.e.~massive planets with short periods, small semi-major axes, high
eccentricities) have been also discovered
\citep[e.g., HAT-P-2b, XO-3b;][]{Bakos2007a,Johns-Krull2008a}. The high
eccentricities are somewhat surprising, as hot Jupiters with short
periods are generally expected to be  circularized in timescales
shorter than the lifetime of the system if the parameter $Q$,
inversely proportional to the planet's tidal dissipation rate, is
assumed to be similar to that inferred for Jupiter
\citep{GoldreichSoter1966,Rasio1996}.

By virtue of their small semi-major axes and high eccentricities, the
longitude of periastron $\omega$ of some of the newly discovered
systems are expected to precess due to General Relativistic (GR)
effects at rates of degrees per century. This is orders of magnitude
larger than the same effect observed in Mercury in our Solar System
($43\arcsec$/century), which offered one of the cornerstone tests of
GR.
Furthermore, the massive, close-in eccentric planets induce
significant reflex motion of the host star, therefore enhancing the
detectability of the precession directly via radial velocities.

In this work we explore the observability of the precession of the
longitude of periastron with the magnitude expected from GR in
exoplanets using radial velocity and transit timing observations. We
also consider in this work the periastra precession due to planetary
perturbers and tidal deformations, which can have contributions
comparable or greater than that of GR.
Previous works \citep{Miralda-Escude2002a,Heyl2007a} have explored
some aspects of the work presented here in the context of using timing
observations to detect terrestrial mass planets. We refer the reader
to independent work by \cite{Pal2008b} that also explores the
measurable effects of the periastra precession induced by GR.

\section{Expected Precession of Periastra}
\label{sec:prec}

Before continuing let us fix our notation. In what follows $a$ will
denote the semi-major axis of the Keplerian orbit of the planet-star
separation, $e$ its eccentricity, $P$ its
period, $\omega$ its longitude of periastron, $\mstar$ and $\rstar$
the mass and radius of the host star, respectively, and $n\equiv
(GM_{tot}/a^3)^{1/2}$ is the Keplerian mean motion (orbital angular
frequency), where $G$ is Newton's gravitational constant and $M_{\rm
tot}=\mstar+\mpl$ with $\mpl$ the mass of the planet.  The reflex
motion of the host star is characterized in a similar manner.  In this
section we detail the expected mechanisms that will cause a precession
in the value of $\omega$.

\subsection{General Relativistic Precession}

One of the most well-known consequences of General Relativity is that
orbits in a Schwarzschild metric are no longer closed as is the case
for the Kepler problem in Newtonian mechanics. The rate of precession
of the longitude of periastron due to GR is given to leading
post-Newtonian order by
\be
\dot{\omega}_{\rm GR} = \frac{3 G \mstar }{a c^2 (1-e^2)}n
\ee
\citep[see any general relativity textbook, e.g.][for a
derivation]{MTW}.  With $a$ expressed in astronomical units, $P$ in
days and the mass of the host star $M_*$ in solar units,
$\dot{\omega}_{\rm GR}$ is given in units of degrees per century by
\be
\dot{\omega}_{\rm GR} = \frac{7.78}{(1-e^2)} 
                      \left(\frac{\mstar}{M_\odot}\right) 
		      \left(\frac{a}{0.05\mbox{AU}}\right)^{-1}
		      \left(\frac{P}{\mbox{day}}\right)^{-1}		      
	\,\,\,\,[^\circ/\mbox{century}].
\label{eq:wdotGR}
\ee
We use equation~\ref{eq:wdotGR} to estimate the expected precession
for currently known exoplanets as listed in the online California \&
Carnegie Catalog of Exoplanets \citep[][version Nov 7
2007]{ButlerCat2006}. We list in Table~\ref{tab:prec} all exoplanets
that have $e>0.1$ and that have $\dot{\omega}_{\rm GR} > 1
^\circ/\mbox{century}$\footnote{The magnitude of all the effects
discussed in what follows where \dotw would manifest itself are increasing
functions of $e$.  Systems with low $e$ are therefore not relevant
from the point of view of detecting GR effects. \citet{Heyl2007a}
presents in their Figure~4 an estimate of \dotw for all systems in
\citet{ButlerCat2006} and note the four systems with higher
values. None of those are in Table~\ref{tab:prec} because they all
have $e< 0.03$.}.  From left to right, the columns in
Table~\ref{tab:prec} record the name of the exoplanet, its semi-major
axis, its eccentricity, the mass $\mstar$ of the host star, the
period, the velocity amplitude\footnote{This is often referred to as the
semi-amplitude in the exoplanet literature. We choose to use simply
amplitude in this paper in order to agree with the widespread usage in
the physical sciences for the multiplicative factor in a simple
harmonic oscillator.}, the longitude of periastron $\omega$, the
calculated $\dot{\omega}_{\rm GR} $ and finally a flag that is 1 if
the planet is transiting its host star and 0 otherwise.
The condition on $e$ is in order to consider only systems where GR
effects can be constrained with sufficient confidence.  Additionally,
we restrict ourselves to stars which have evidence currently for a
single exoplanet in order to avoid the complications arising from the
precession of $\omega$ induced by other planets (see
below)\footnote{\citet{Adams2006a,Adams2006b} studied the effects of
secular interactions in multiple-planet systems including the effects
of GR. They show that GR can have significant effects on secular
perturbations for systems with favorable characteristics.}.  We have
added to the exoplanets listed by \citet{ButlerCat2006} the recently
discovered XO-3b
\citep{Johns-Krull2008a}, which has the largest predicted
$\dot{\omega}_{\rm GR}$ of all exoplanets listed in Table~\ref{tab:prec}.

\begin{deluxetable}{lccccrrcc}
\tablecaption{GR precession of exoplanets with $e>0.1$ and GR 
Precession Rates $>1^\circ$/century}
\tabletypesize{\footnotesize}
\tablewidth{0pt}
\tablehead{
\colhead{Name} & \colhead{$a$} & \colhead{$e$} 
& \colhead{$\mstar$ } & \colhead{Period}
& \colhead{K} & \colhead{$\omega$}
& \colhead{$\dot{\omega}_{\rm GR} $} & \colhead{TEP?}\\
 \colhead{} & \colhead{(AU)} & \colhead{} 
& \colhead{$(M_\odot)$} & \colhead{(days)}
& \colhead{(m sec$^{-1}$)} & \colhead{(deg)} 
& \colhead{$^\circ$/century} & \colhead{}
}
\startdata
HD49674    b & 0.058     & 0.290     & 1.060     & 4.944     & 13.7    & 283.0   & 1.576     & 0 \\
HD88133    b & 0.047     & 0.133     & 1.200     & 3.416     & 36.1    & 349.0   & 2.958     & 0 \\
GJ 436     b & 0.028     & 0.159     & 0.410     & 2.644     & 18.7    & 339.0   & 2.234     & 1 \\
HD118203   b & 0.070     & 0.309     & 1.230     & 6.133     & 217.0   & 155.7   & 1.231     & 0 \\
HAT-P-2    b & 0.069     & 0.507     & 1.350     & 5.633     & 884.0   & 184.6   & 1.836     & 1 \\
HD185269   b & 0.077     & 0.296     & 1.280     & 6.838     & 90.7    & 172.0   & 1.046     & 0 \\
XO-3       b & 0.048     & 0.260     & 1.410     & 3.192     & 1471.0 & -15.4    & 3.886     & 1 \\
\enddata
\label{tab:prec}
\end{deluxetable}

As can be seen in Table~\ref{tab:prec}, seven systems have
$\dot{\omega}_{\rm GR} > 1 ^\circ/\mbox{century}$, with 3 of them
being transiting systems. In timescales of a few tens of years, the
longitude of periastron of the systems is expected to shift in these
systems by $\delta \omega \gtrsim 0.5^\circ$, a change that, as we
will show below, may produce detectable effects. GR effects are not
the only mechanisms that can cause a shift in $\omega$ though, so we
now turn our attention to additional mechanisms. 

\subsection{Stellar Quadrupole, Tides and Additional Planets}
\label{ssec:add_prec}

Besides the GR precession discussed above, $\omega$ can precess due to
the presence of additional effects. \citet{Miralda-Escude2002a}
discussed the observability, using the duration of transits and the
period between transits, of changes in $\omega$ caused by a stellar
quadrupole moment and perturbations from other planets. These effects,
some of which were discussed using more accurate calculations by
\citet{Heyl2007a}, may additionally cause a precession of the orbital plane. 
Additionally, tidal deformations induced on the star and the planet
can also produce a secular change on $\omega$, an effect not
considered in the studies mentioned above. The effect of apsidal
motions induced by tidal deformations is a well known effect in
eclipsing binaries
\citep{Sterne1939a,Quataert1996a,Smeyers2001a}, and was included in
the analysis of the planetary system around HD~83443 by
\citet{Wu2002a}. Tidal deformations can produce a significant amount
of precession in close-in exoplanets and should therefore be taken
into account.

The precession caused by a stellar quadrupole moment is given 
to second order in $e$ and first order in $(R_*/a)^2$ by
\be
	\dot{\omega}_{\rm quad} \approx \frac{3 J_2 \rstar^2}{2a^2} n,
\ee
\noindent where $J_2$ is the quadrupole moment
\citep{MurrayDermott}. In units of degree/century this expression
reads
\be
	\dot{\omega}_{\rm quad} \approx 0.17
	\left(\frac{P}{\mbox{day}}\right)^{-1}
	\left(\frac{J_2}{10^{-6}}\right)
	\left(\frac{\rstar}{R_\odot}\right)^2
	\left(\frac{a}{0.05\mbox{AU}}\right)^{-2}
	\,\,\,\,[^\circ/\mbox{century}].
\ee
It is clear from this equation that for values of $J_2\lesssim
10^{-6}$ similar to that of the Sun \citep{Pireaux2003a} the value of
$\dot{\omega}_{\rm quad}$ is smaller than the value of
$\dot{\omega}$ expected from GR \citep[see
also][]{Miralda-Escude2002a}. We will therefore assume in what follows
that $\dot{\omega}_{\rm quad}$ is always negligible in comparison with
$\dot{\omega}_{\rm GR}$.

The tidal deformations induced on the star and the planet by each
other will lead to a change in the longitude of periastron which is
given, under the approximation that the objects can instantaneously
adjust their equilibrium shapes to the tidal force and considering
up to second order harmonic distortions, by
\be
    \dot{\omega}_{\rm tide} \approx \frac{15 f(e)}{a^5}
    \left( \frac{k_{2,s}\mpl \rstar^5}{\mstar}	+ 
    \frac{k_{2,p}\mstar \rpl^5}{\mpl}
    \right) n,
\label{eq:tideN}
\ee
\noindent where $f(e)\equiv(1-e^2)^{-5}[1+(3/2)e^2+(1/8)e^4]$, 
and $k_{2,s}, k_{2,p}$ are the apsidal motion constants for the star
and planet respectively, which depend on the mass concentration of the
tidally deformed bodies \citep{Sterne1939a}. For stars we expect
$k_{2,s} \lesssim 0.01$ \citep{Claret1992a}, while for giant planets
we expect $k_{2,p} \approx 0.25$ if we assume that their structure can
be roughly described by a polytrope of index $n
\approx 1$ \citep{Hubbard1984a}. For the extreme case of a sphere of uniform mass
density, the apsidal motion constant takes the value $k_2=0.75$
\citep[e.g.,][]{Smeyers2001a}.
We see from Equation~\ref{eq:tideN} that for close-in hot Jupiters the
term containing $k_{2,p}$ will dominate, and that the effect of tides
on $\omega$ increases very rapidly with decreasing $a$. 
In units of degree/century equation~\ref{eq:tideN} gives
\be
\dot{\omega}_{\rm tide} \approx
	  1.6 f(e) {\cal T}
	  \left(\frac{P}{\mbox{day}}\right)^{-1}
	\left(\frac{k_{2,p}}{0.1}\right)
	\left(\frac{a}{0.05\mbox{AU}}\right)^{-5}
	\left(\frac{\rpl}{R_J}\right)^5
	\left(\frac{M_J}{\mpl}\right)
	\left(\frac{\mstar}{M_\odot}\right)	
\,\,\,\,[^\circ/\mbox{century}],
\ee
\noindent where we have introduced 
${\cal T} \equiv 1 + (\rstar/\rpl)^5(\mpl/\mstar)^2(k_{2,s}/k_{2,p})$, which
is $\approx 1$ for the case of a close-in Jupiter.
Assuming that $k_{2,p} \sim 0.1$, $e\lesssim 0.5$, $\mpl \sim M_{J}$,
$\mstar \sim M_\odot$, $\rstar \sim R_\odot$ and $\rpl \sim R_J$
it follows that $\dot{\omega}_{\rm tide}$ is of comparable
magnitude as $\dot{\omega}_{\rm GR}$. 

The precession of the periastra caused by a second planet, which we
dub a ``perturber'', is given to first order in $e$  and lowest order
in $(a/a_2)$ by 
\be
\dot{\omega}_{\rm perturber} \approx \frac{3M_2 a^3}{4 \mstar a_2^3}n
\label{eq:wdot_pert}
\ee
\citep{MurrayDermott,Miralda-Escude2002a}, where $M_2$ is the
mass of the second planet and $a_2$ the semi-major axis of its
orbit. In terms of deg/century this expression reads
\be
   \dot{\omega}_{\rm perturber} \approx 29.6
   \left(\frac{P}{\mbox{day}}\right)^{-1}
   \left(\frac{a}{a_2}\right)^{3}
   \left(\frac{\mstar}{M_\odot}\right)^{-1}	
   \left(\frac{M_2}{M_\oplus}\right)   
   \,\,\,\,[^\circ/\mbox{century}].
\ee

For a perturber with $a_2 = 2a$ and and a mass similar to Earth
orbiting a solar-mass star, we get that $\dot{\omega}_{\rm perturber}
\sim 3\times 10^{-7}n$ or $\dot{\omega}_{\rm perturber}
\sim 0.7 $ deg/century for a $P=5$ days planet. This is comparable
to the precession expected from GR and therefore any detected
precession of the longitude of periastron will be that of GR plus the
possible addition of any perturber planet present in the system and
the effects of tidal deformations (and generally a negligible
contribution from the stellar quadrupole). 

In what follows we will discuss the observability of changes in
$\omega$ using radial velocity and transit observations. As just
shown, any precession is expected to arise by GR, the effect of
additional planets or tidal deformations. The discussion that follows
addresses the detectability of changes in $\omega$ independent of
their origin.

\section{Observability of Periastra Precession in Extra-Solar Planets}

\subsection{Radial Velocities}
\label{subsec:rv}

The radial velocity of a star including the reflex motion due to a
planetary component is given by
\be
v_r(t) = v_0 + K[\cos(\omega + f(t-t_0)) + e\cos(\omega)],
\label{eq:rv}
\ee
where $v_0$ is the systemic velocity,  $t_0$ the time
coordinate zeropoint, $f$ the true anomaly and $K$ is the velocity
amplitude which is related to the orbital elements and the masses
by
\be
	K = \left( \frac{2\pi G}{P}\right)^{1/3} \frac{\mpl \sin i}{M_{\rm tot}},
\ee

\noindent where $i$ is the orbit inclination. Fitting for the observed
radial velocities of a star will give then direct estimates of $v_0,
t_0, K, e, P$ and $\omega$.

Our aim in this section is to determine if $\omega$ can be constrained
tightly enough in timescales of tens of years or less in order to
detect changes in $\omega$ of the magnitude produced by GR in those
time-spans.
In order to do this we have simulated data and then fit it with a
model of the form given by equation~\ref{eq:rv} a total of 1000
times. We then recover the best-fit values of $\omega$ in all
simulations and use that to estimate the probability distribution
$\phi(\omega)$ expected under given assumptions.

The systemic velocity $v_0$ and $t_0$ are just zero-points that we set
to 0 in all our simulations, where we also set the time units such
that $P=1$ (note though that we do fit for all these quantities so
that their effect on the fit propagates to the uncertainties of
$\omega$).  By trying several values of $\omega$ we have verified that
the probability distributions recovered do not depend strongly on the
particular value of $\omega$, which we therefore fix for all
simulations at an arbitrary value $\omega_0=135^\circ$.  This leaves
us with just two parameters to vary, namely $e$ and $K$.

Given $e$ and $K$ we generate $N_{\rm obs}$ data-points with times
$t_i$ uniformly distributed\footnote{We ignore in
our simulations the Rossiter-McLaughlin effect for the case of
transiting planets
\citep{Rossiter1924a,McLaughlin1924a,Queloz2000a}.}
within a period and then we add to each
time a random number of periods between 0 and 20. For transiting
exoplanets the observations would have to be taken uniformly in time
intervals excluding the transit. We then get the observed radial
velocity from equation~\ref{eq:rv} as

\be
	v_{r}(t_i) = K[\cos(\omega_0 + f(t_i)) + e\cos(\omega_0)] + 
	G(0,\sigma_{\rm tot})
	\label{eq:vr}
\ee

\noindent where $G(0,\sigma_{\rm tot})$ is a random Gaussian deviate
with mean 0 and standard deviation $\sigma_{\rm tot}$. The latter
quantity is obtained as $\sigma_{\rm tot}^2 = \sigma_{\rm obs}^2 +
\sigma_{\rm jitter}^2$, where $\sigma_{\rm obs}$ is the random
uncertainty for each measurement and $\sigma_{\rm jitter}$ is the
noise arising from stellar jitter. Given the form of
equation~\ref{eq:rv} a Fisher matrix analysis implies that the
uncertainty in the longitude of periastron, $\sigma_\omega$ satisfies
the following scaling

\be
	\sigma_\omega \propto N_{\rm obs}^{-1/2}\sigma_{\rm tot} K^{-1}.
	\label{eq:scaling}
\ee

This scaling allows us to perform a set of fiducial simulations for
several values of $e$ and use the results to scale to parameters
relevant to a given situation of interest. 
We note that the simulations we performed verify that the scaling
inferred from a Fisher matrix analysis is accurate.
In order to measure the longitude of periastron $\omega$ with a
reasonable degree of certainty the system clearly needs to have a
significant amount of eccentricity. We restrict ourselves to systems
with $e\ge0.1$ and we simulate systems with
$e=0.1,0.2,0.3,0.4,0.5,0.6$.

For $\sigma_{\rm obs}$ we assume a typical high-precision measurement
with $\sigma_{\rm obs}=2$ m sec$^{-1}$. For the stellar jitter, we
perform our fiducial simulations for a typical jitter of $\sigma_{\rm
jitter} = 4$ m sec$^{-1}$ \citep{wright2005,ButlerCat2006}.
The limitations imposed by active stars and/or different precision on
the radial velocity measurements on the recovery of \dotw can be
explored by using the scaling of $\sigma_\omega$ with higher assumed
values of $\sigma_{\rm jitter}$ and/or $\sigma_{\rm obs}$.
Even though Equation~\ref{eq:scaling} renders multiple values of $K$
redundant, we choose to present results for two values of $K$ for
illustrative purposes, namely $K=100$ m sec$^{-1}$ and $K=1000$ m
sec$^{-1}$. The former corresponds roughly to Jupiter-mass exoplanets
and is fairly representative of currently known systems
\citep{ButlerCat2006}\footnote{See also \url{http://www.exoplanet.eu}.}, 
while the latter corresponds to super-massive planets which as we will
see are the class of systems which would allow the detection of
$\dot{\omega}_{\rm GR}$ with radial velocities.
Finally, we do our fiducial simulations for $N_{\rm obs} =100$, a
value not atypical for well-sampled radial velocity curves available
today.

\begin{deluxetable}{cccc}
\tablecaption{Results of Radial Velocity Curves Fit Simulations for 
$N_{\rm obs}=100$, $\sigma_{\rm obs}=2$ m/sec and 
$\sigma_{\rm jitter}=4$ m/sec}
\tabletypesize{\footnotesize}
\tablewidth{0pt}
\tablehead{
\colhead{$K$} & 
\colhead{$e$ } & 
\colhead{$\sigma_\omega$} & 
\colhead{$\alpha_{20}$} \\
 \colhead{(m sec$^{-1}$)} 
& \colhead{} & 
\colhead{(deg)} & 
\colhead{(deg/century)}
}
\startdata
  100 &  0.10 &  3.77 & 80.010\\
  100 &  0.20 &  1.85 & 39.204\\
  100 &  0.30 &  1.29 & 27.457\\
  100 &  0.40 &  1.05 & 22.249\\
  100 &  0.50 &  0.95 & 20.253\\
  100 &  0.60 &  0.86 & 18.159\\
 1000 &  0.10 &  0.38 & 8.059\\
 1000 &  0.20 &  0.19 & 4.106\\
 1000 &  0.30 &  0.13 & 2.768\\
 1000 &  0.40 &  0.10 & 2.182\\
 1000 &  0.50 &  0.09 & 1.959\\
 1000 &  0.60 &  0.08 & 1.790\\
\enddata
\label{tab:sim}
\end{deluxetable}

The results of the simulations are summarized in Table~\ref{tab:sim}.
From left to right, the columns in this Table record the assumed
radial velocity amplitude $K$, the system eccentricity $e$, the
expected uncertainty in the longitude of periastron $\sigma_\omega$
for the simulated system, and finally $\alpha_{20}$, which we define
to be the value of $\dot{\omega}$ that would be necessary in order to
achieve a 3$\sigma$ detection in the simulated systems in a time-span
of 20 years when measuring $\omega$ in two epochs, each having $N_{\rm
obs}$ observations. We note that we have performed Shapiro-Wilk
normality tests in the recovered $\omega$ distributions. We found that
all of them are consistent with normality and thus we are justified in
using the dispersion $\sigma_w$ to derive confidence levels.

\begin{figure}
\epsscale{0.6}
\plotone{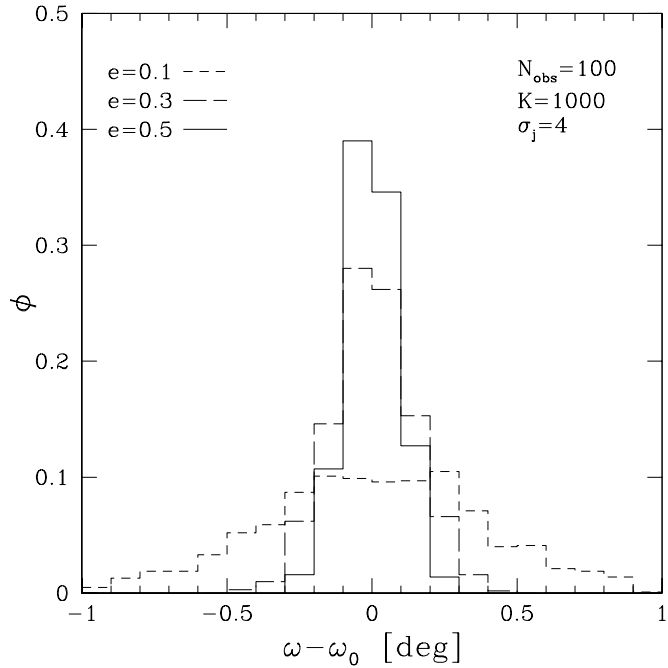}
\caption{
	Distribution $\phi$ of recovered angles of periastron
	$\omega-\omega_0$, where $\omega_0$ is the input angle, for the
	simulation with $N_{\rm obs}=100$, $K=1000$ m sec$^{-1}$ and
	$\sigma_j=4$ m sec$^{-1}$. The distributions are shown for
	eccentricities $e=0.1,0.3,0.5$.
\label{fig:rv_eg}
\vspace{0.2cm}
}
\end{figure}

\begin{figure}
\epsscale{0.6}
\plotone{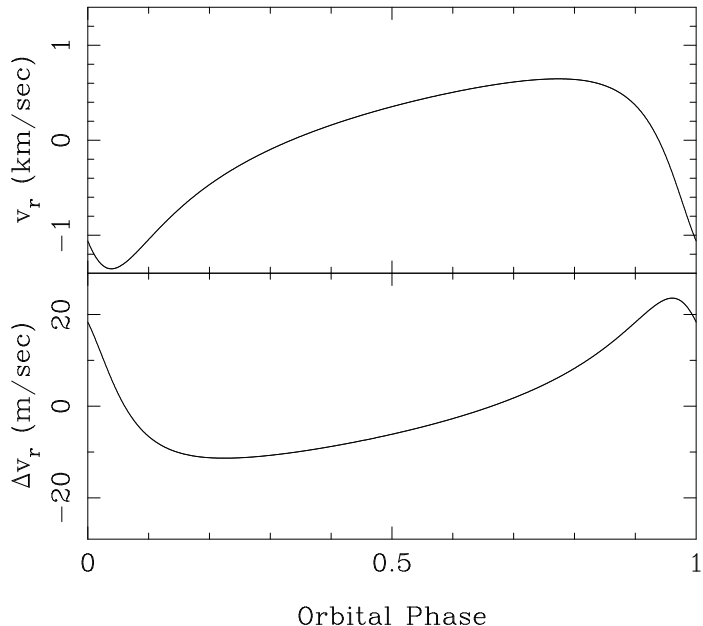}
\caption{
	({\it Top}\/) Radial velocity curve as a function of orbital phase
	for a system with $K=1000$ m sec$^{-1}$, $e=0.5$ and
	$\omega=135^\circ$.  ({\it Bottom}\/) Difference between the curve in
	the top panel and a radial velocity curve that is identical except
	in that it has $\omega=136^\circ$.  Note the different $y$-axis
	scales used in the different panels.
	\label{fig:rv_eg2}
	\vspace{0.2cm}
}
\end{figure}

As an example, we show in Figure~\ref{fig:rv_eg} the distributions of
recovered $\omega$ for the case with $N_{\rm obs}=100$, $K=1000$ m
sec$^{-1}$ and $\sigma_j=4$ m sec$^{-1}$ for eccentricities
$e=0.1,0.3,0.5$.
In the upper panel of Figure~\ref{fig:rv_eg2} we show the radial
velocity curve for a system with $K=1000$ m sec$^{-1}$, $e=0.5$ and
$\omega=135^\circ$, while in the lower panel we show the difference
between that curve and a radial velocity curve having
$\omega=136^\circ$. The simulations presented in Table~\ref{tab:sim}
show that this difference can be detected at the 10-$\sigma$ level
($\sigma_\omega \sim 0.1^\circ$) when using $N_{\rm obs}=100$
observations, each having an uncertainty of $\sim 2$ m sec$^{-1}$.
The high level of significance can be achieved in this case thanks to
the assumed low-level of jitter and the super-massive nature of a
system with that semi-velocity amplitude.

In order to validate our simulations against uncertainty estimates
obtained with real data, we have run a simulation with $N_{\rm
obs}=20$, $e=0.517$, $K=1011$ m sec$^{-1}$, $\omega=179.3^\circ$ and $\sigma_j=60$ m
sec$^{-1}$. These observational conditions and orbital parameters are
appropriate for the observations of HAT-P-2 b reported by
\citet{Bakos2007a}. Our simulations for this case return
$\sigma_w=3.8^\circ$, while \citet{Bakos2007a} report $\omega_{\rm
HAT-P-2 b} = 179.3 \pm 3.6^\circ$, in very good agreement with our
estimate. We conclude that our simulations return realistic estimates
of the uncertainties in $\omega$.

The simulations presented in Table~\ref{tab:sim} show that in some
cases the precession of periastron detectable in 20 years,
$\alpha_{20}$, is comparable to the values $\dot{\omega}_{GR}$ of
currently known systems listed in Table~\ref{tab:prec}. For close-in,
eccentric, super-massive planets ($K\sim 1000$ m sec$^{-1}$) about 100
observations per epoch are sufficient, while for Jupiter-mass
exoplanets ($K\sim100$ m sec$^{-1}$) an unrealistically large number
of radial velocity observations, $N_{\rm obs}=10000$ per epoch, would
be needed. Therefore, observations of different epochs of radial
velocities with a time-span of $\sim$ 20 years would detect the
variations in the precession of exoplanet periastra induced by GR in
some currently known super-masive systems. The estimates above assume
a typical level of stellar jitter. For active stars the number of
observations have to be increased in proportion to the dispersion
characterizing the jitter (see Equation~\ref{eq:scaling}). Stellar
activity is therefore an important limitation to detect changes in
$\omega$ using radial velocities. All in all, radial velocity studies
of exoplanets will be generally able to ignore the effects of GR
precession as they will usually be well below a detectable level.

We note in closing that \citet{Miralda-Escude2002a} and 
\citet{Heyl2007a} consider using radial velocities and transit timing 
observations 
in order to measure the small difference between the period observed
in the radial velocities and the period between primary transits.
Both works conclude that this is not a competitive method and so we
will not consider it further here and refer the reader to those works
for details.

\subsection{Duration of Transits}
\label{subsec:D}

As the planetary orbit acquires a significant eccentricity $e$, the
duration of the primary and secondary transits are no longer equal and
acquire a dependence on the longitude of periastron of the system. An
explicit expression for the duration of a transit $D$ in the eccentric
case was derived by \citet[][their equation 7]{Tingley2005} under the
assumption that the distance between the planet and the star does not
change significantly during transit. It is given by
\be
	D = 2Z(\rstar + \rpl)\frac{\sqrt{1-e^2}}{(1+e\cos(f_t))}
	\left(\frac{P}{2\pi G M_{\rm tot}}\right)^{1/3},
\ee

\noindent where 

\be
Z = \sqrt{1-\frac{r_t^2\cos^2 i}{(\rstar+\rpl)^2}} \equiv \sqrt{1-b^2}
\ee

\noindent is a geometrical factor related to the impact parameter 
$b$.  $\rstar$ and $\rpl$ are the radii of the star and planet
respectively, $M_{\rm tot} \equiv \mstar + \mpl$ and $r_t$ and $f_t$
are the radii and true anomaly at the time of transit,
respectively. The latter clearly depends linearly on $\omega$ and we
therefore have $\dot{f}_t = \dot{\omega}$. The logarithmic derivative
of the duration of a transit is

\be
d\ln D/dt = \frac{\dot{\omega} e \sin(f_t)}{(1+e\cos(f_t))}
\left\{1-\frac{b^2}{1-b^2}\right\}.
\label{eq:dlnD}
\ee

\begin{figure}
\epsscale{0.6}
\plotone{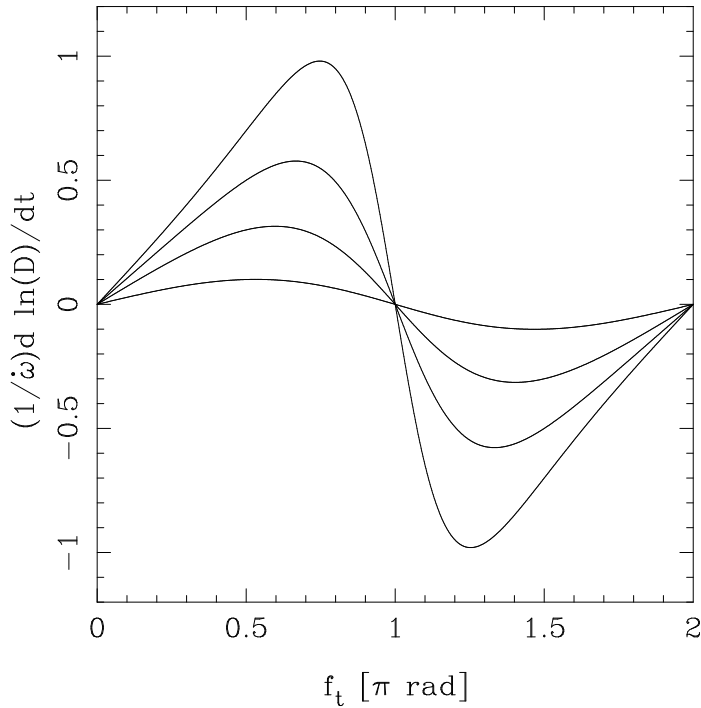}
\caption{
	$(1/\dot{\omega}) d\ln D/dt$ as a function of the true anomaly
	at the time of transit $f_t$ for a central transit
	($p=0$). The different curves are for difference
	eccentricities $e=0.1,0.3,0.5,0.7$, with higher $e$ giving
	higher values of $|(1/\dot{\omega}) d\ln D/dt|$
	\label{fig:dlnDdt}. The extrema in this figure are at values
	of the true anomaly at the time of transit of $f_t=
	\pm\arccos(-e)$.  }
\end{figure}

In Figure~\ref{fig:dlnDdt} we show the quantity $(1/\dot{\omega}) d\ln
D/dt$ for a central transit ($b=0$) and for $e=0.1,0.3,0.5,0.7$, with
higher $e$ giving higher values of $|(1/\dot{\omega}) d\ln D/dt|$. The
change in the duration of an eclipse for a small change in $\omega$
given by $\dot{\omega}\delta t$ is simply $\delta D \sim D
\dot{\omega}\delta t \,\,[\dot{\omega}^{-1}d\ln D/dt] $.  For
$\dot{\omega}\delta t \sim 0.5\times10^{-2}$ rad, appropriate for the
expected change in $\omega$ for $\dotw \sim 3$ deg/century over 10
years, we have that $\delta D \sim 0.075D \times 10^{-2}$ for $e \sim
0.3$, which translates into $\delta D\sim 10$ sec for a transit
duration of $D \sim 0.15$ day which is typical for the sytems that we
explore in this work.

A very interesting feature of Equation~\ref{eq:dlnD} is its dependence
on the impact parameter $b$. First, $d\ln D/dt$ vanishes for
$b=1/\sqrt{2}$. This behavior is possible due to two competing effects
which cancel out exactly for that value of $b$: an increase/decrease
in the star-planet separation causes both and decrease/increase in the
path-length of the planet across the disk of the star and a
corresponding decrease/increase on the velocity across it, which
implies a slower/faster crossing-time. Secondly, for systems with
values of $b$ close to 1 (i.e., near-grazing systems), the value of
$d\ln D/dt$ increases greatly. We will come back to near-grazing
systems below; in what immediately follows we will quantify the
accuracy to which we can determine the transit duration $D$.

If the times of beginning of ingress and end of egress are denoted by
$t_i$ and $t_e$ respectively, then the duration of a transit is given
by $D=t_e-t_i$ and the uncertainty in the duration is
$\sigma_D^2=\sigma_{t_i}^2 + \sigma_{t_e}^2$ (assuming no correlation
between $t_i$ and $t_e$). If the ingress has a
duration of $\Delta t_i$ (i.e.~the time from first contact to second
contact) then a linear approximation of the flux during this time can 
be written as $F(t) =
F_0(1-(t-t_i)(\rpl/\rstar)^2/\Delta t_i)$, where $F_0$ is the
out-of-transit stellar flux and we ignore the effects of limb
darkening. If the light curve is being sampled at a rate $\Gamma$ per
unit time then we have $N= \Delta t_i \Gamma$ photometric measurements
during the egress. If each photometric measurement has a fractional
precision $\sigma_{\rm ph}$, and assuming $\rpl$ and $\rstar$ are
known, a least-squares fit to the photometric series will allow the
determination of $t_i$ with an uncertainty $\sigma_{t_i} = \sigma_{\rm
ph} \Delta t_i (\rstar/\rpl)^2 N^{-1/2} =
\sigma_{\rm ph} (\Delta t_i/\Gamma)^{1/2}(\rstar/\rpl)^2$. 
The uncertainty in the time of egress is given by a similar expression
replacing $\Delta t_i$ by $\Delta t_e$.

If we assume that the ingress and egress times are equal\footnote{The
ingress and egress times are not equal in general for eccentric
systems, see Equation~7 in \citet{Ford2008a}.  This small difference
has no significant effect on the uncertainty estimates dealt with
here.},  and insert explicit expressions for $\Delta t_i \approx
\Delta t_e$ and $D$, we get that for a central transit
\citep{Ford2008a}:

\be
\frac{\sigma_D}{D} \approx \frac{40.7 \sigma_{\rm ph}\sqrt{1+e\cos f_t}}{(1-e^2)^{1/4}}
	\sqrt{\frac{2}{N_{tr}\Gamma}}
	\left( \frac{\rpl}{R_\oplus}\right)^{-3/2}
	\left( \frac{\rstar}{R_\odot}\right)
	\left( \frac{\mstar}{M_\odot}\right)^{1/6}	
	\left( \frac{P}{{\rm yr}}\right)^{-1/6}
	(1+\rpl/\rstar)^{-3/2},	
	\label{eq:sigmaD}
\ee

\noindent where $N_{tr}$ is the number of transits observed, $\Gamma$
is to be expressed in units of min$^{-1}$, and we have neglected
factors including $(1+\mu)$, where $\mu=\mpl/\mstar$. We note for later reference that the
uncertainty in the central transit time $t_c \equiv 0.5(t_e + t_i)$ is
$\sigma_ c \approx \sigma_D/\sqrt{2}$.

The highest photometric precision in a transit light curve has been
achieved with {\it HST} \citep{Brown2001a,Pont2007a} with $\sigma_{\rm ph}
\sim 10^{-4}$.  For parameters appropriate to HD~209458
Equation~\ref{eq:sigmaD} gives $\sigma_D \sim 5$ sec. The
uncertainties in the central time $\sigma_c \sim \sigma_D$ reported in
\citet{Brown2001a} are of the same order and we therefore deem
Equation~\ref{eq:sigmaD} to be a reasonable estimate of the expected
uncertainties\footnote{We note that \citet{Brown2001a} warn about the
presence of systematic effects not accounted for in the Poisson
uncertainty estimate that could be of the same order as $\sigma_D$ for
their observations.}. We will take this to be the highest precision
currently possible for facilities such as {\it HST} where only a few
transits are typically observed.

The {\it Kepler} mission \citep{Borucki2003a,Basri2005a} will observe $\sim$100,000 stars
continuously with $\sigma_{\rm ph} \sim 4 \times 10^{-4}$ for a one
minute exposure of $V=12$ solar-like star and expects to find
a large number of close-in ``hot Jupiters''. The precision attainable in one
year for a 1-minute sampling of a Jupiter-mass system with $P=5$ days
orbiting a $V=12$ solar-type star is $\sigma_D/D \sim
1.1\times10^{-4}$ or $\sigma_D \sim 1.5$ sec for a 0.15 days
transit duration.

{\it Kepler} should therefore be capable of detecting changes in the
transit duration $D$ due to GR within its mission for some eccentric,
close-in systems, and certainly when coupled to follow-up
determinations of $D$ with a facility delivering a precision similar
to what {\it HST} can achieve.
Of particular interest will be systems that are near-grazing, due to
the factor of $(1-b^2)^{-1}$ in Equation~\ref{eq:dlnD}. If eccentric,
close-in, near-grazing systems are found in the {\it Kepler} CCDs,
they will be subject to significant changes in $D$ which should be
observable. For example, a system with $e=0.4$, $b=0.85$, $f_t=\pi/2$
and $\dotw \sim 3$ deg/century would have a change in $D$ of
$\sim 9$ sec during {\it Kepler's} lifetime for $D=0.075$ days, a
change which would be detectable. Of course, as $b$ approaches one, $D$
will tend to zero and the assumptions leading to
Equation~\ref{eq:sigmaD} will break down and make the estimated
uncertainty optimistic, both effects which will counter the
corresponding increase of $d\ln D/dt$.

\subsection{Period Between Transits}
\label{subsec:dPdt}

As already noted by \citet{Miralda-Escude2002a} and \citet{Heyl2007a},
as the longitude of periastron changes, the period between transits
$P_t$ will change as well. Periods are the quantities that are
measured with the greatest precision, usually with uncertainties
on the order of seconds from ground-based observations
(e.g., the average uncertainty for HATNet planets discovered to date is
4.5 seconds).

To first order in $e$ the derivative of the transit period is given by

\be
\dot{P}_t = 4\pi e\left(\frac{\dotw}{n}\right)^2\sin (M_t)
\ee

\noindent where $M_t$ is the mean anomaly at transit 
\citep{Miralda-Escude2002a} and is related to the true anomaly at 
transit $f_t$ to first order in $e$ by $M_t = f_t - 2e\sin f_t$.  For
$e=0.1$, a 5 day period, and $\dotw=3$ deg/century, the root mean
square value of $dP_t/dt$ over all possible $M_t$ values is $\sim
10^{-12}$, which translates into a period change of $\sim 2 \times
10^{-4}$ sec in 10 years.  The period can be determined to a precision
$\sim \sigma_c N_{tr}^{-3/2}$ or $\sigma_D N_{tr}^{-1}2^{-1/2}$ using
the expression for $\sigma_D$ above\footnote{The scaling $N^{-3/2}$
follows from describing the central transit times as $t_i = t_0 + Pi$,
with $i=1, \ldots, N_{tr}$ and determining $P$ using a $\chi^2$
fit. The variance in the derived $P$ is \citep[see,
e.g.,][]{Gould2003a} $\sigma^2_P \approx (3/N^3)\sigma_c^2 + {\cal
O}(N^{-4})$.}.
Assuming the parameters for {\it Kepler} as above ($V=12$ star, $P=5$
day period) the precision achievable during 1 year is $\sim 0.013$
sec.

Based on the numbers above we conclude that measuring significant
changes in the transiting period in $\lesssim 10 $ years timescales is
not feasible.  Our conclusions are in broad agreement with the
analysis presented in \citet{Miralda-Escude2002a} and
\citet{Heyl2007a}, who conclude that thousands of transits need to be
observed with high precision in order to detect significant variations
in $P_t$. As there is no existing or planned facility that will allow
to observe this amount of transits with the required photometric
precision we conclude that measurements of $\dot{P}_t$ will not be
significantly affected by changes in $\omega$ of the magnitudes expected to
arise from GR or from the secular changes due to a perturber.

\subsection{Time Between Primary and Secondary Transit}
\label{subsec:Deltat}

In the case where the exoplanet is transiting it may be possible to
observe not only the primary transit, i.e.~the transit where the
exoplanet obscures the host star, but also the occultation, when the
host star blocks thermal emission and reflected light from the
exoplanet
\citep[e.g.,][]{Charb2005}.
If the time of the primary eclipse is given by $t_1$ and that of the
secondary by $t_2$, the time difference between the two as compared to
half a period $P$, $\Delta t \equiv t_2 - t_1 - 0.5P$, depends mostly
on the eccentricity $e$ and the angle of periastron $\omega$. Indeed,
an accurate expression that neglects terms proportional to $\cot^2 i$
where $i$ is the inclination angle and is therefore exact for central
transits, is given by
\be
	\Delta t= \frac{P}{\pi}\left(\frac{e\cos(\omega)
	\sqrt{1-e^2}}{(1-(e\sin\omega)^2)} + 
	\arctan\left( \frac{e\cos\omega}{\sqrt{1-e^2}}\right)\right)
\label{eq:sterne}
\ee

\noindent \citep{Sterne1940}. Combined with radial velocities this time
difference offers an additional constrain on $e$ and $\omega$, and in
principle a measurement of $\Delta t$ combined with a measurement of
the difference in the duration of the secondary and primary eclipses
can be used to solve for $e$ and $\omega$ directly \citep[see
discussion in][]{Charb2005}.  We note that \citet[][their
\S4.2]{Heyl2007a} also consider secondary transit timings as a means
to measure changes in $\omega$. While their discussion is based on
first order expansions in $e$ instead of using the exact expression
above and is phrased in different terms, it is based ultimately on the
same measurable quantity we discuss here. 
\footnote{\citet{Heyl2007a} consider the time difference between 
successive primary transits ($\Delta t_0$) and successive secondary
($\Delta t_\pi$) transits, following their notation.  This difference
can be expressed as $\Delta t_0 - \Delta t_\pi \approx -(d \Delta t /
d\omega) \delta \omega$, where $\Delta t$ is the quantity defined in
Equation~\ref{eq:sterne} and $\delta\omega$ is the change in $\omega$
in one orbit.  }

Equation~\ref{eq:sterne} does not include light travel time
contributions, i.e.  it neglects the time it takes for light to travel
accross the system. This time is given for a central transit by

\be
\Delta_{t,LT} = \frac{2a(1-e^2)}{c[1-(e\cos f_t)^2]},
\label{eq:LT}
\ee

\noindent where $f_t$ is the true anomaly at the time of primary transit.
In this section we will be interested in changes in $\Delta t$ due to
changes in $\omega$. It is easy to see from the expressions above that
$d\Delta_{t,LT} / d\omega \ll d\Delta t / d\omega$ (ignoring the
points where they are both zero). Therefore, changes in the time
difference between primary and secondary transits will be dominated by
changes in $\Delta t$ and we can safely ignore light travel time
effects in what follows.

\begin{figure}
\epsscale{0.6}
\plotone{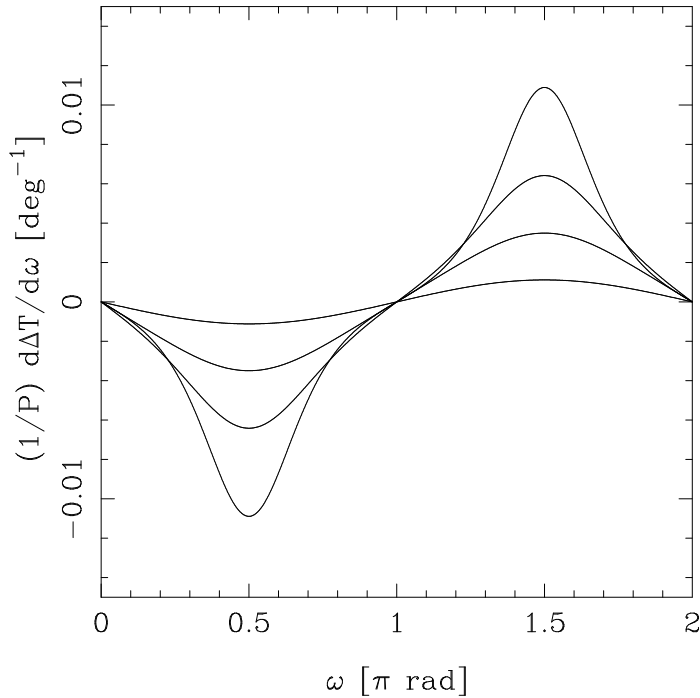}
\caption{
	$(1/P) d\Delta t / d\omega$ as a function of $\omega$. The
	different curves are for different eccentricities
	$e=0.1,0.3,0.5,0.7$, with higher $e$ giving larger extrema of
	$|(1/P) d\Delta t / d\omega|$.  \label{fig:dDt} }
\end{figure}

The precession of periastra caused by GR will change $\omega$
systematically, while leaving $e$ unchanged. Therefore $\Delta t$ has
the potential of offering a direct measurement of changes in $\omega$.
We use Equation~\ref{eq:sterne} to calculate $(1/P) d\Delta t /
d\omega$ as a function of $\omega$. The result is shown in in
Figure~\ref{fig:dDt} for $e=0.1,0.3,0.5,0.7$, with higher $e$ yielding
larger extrema of $|(1/P) d\Delta t / d\omega|$.

If $\omega \approx \pi/2$ or $\approx 3\pi/2$, the changes $\Delta t$ in
a system with $P=4$ days for 0.2 degrees of precession would be large,
on the order of 8 min for $e\sim 0.5$.  Published observations of
secondary eclipses with {\it Spitzer} constrain $t_2$ to within
$\sim 80$ sec \citep{Deming2006a,Harrington2007a,Charbonneau2008a}.
Uncertainties in $t_2$ will dominate the uncertainty in $\Delta_t$,
so we assume $\sigma_{\Delta_t} \approx \sigma_{t_2} \sim
80N_{tr}^{-3/2}$ sec, where $N_{tr}$ is the number of secondary
transits observed.

Determing $\Delta t$ at two epochs separated by 10 years and observing
$N_{tr} \sim 1$ at each epoch with a facility like {\it Spitzer}
would detect the changes in $\Delta t$ due to GR with a significance
$\gtrsim 4 \sigma$ in an eccentric system that precesses $\gtrsim$ 0.2
deg in a decade if they have $\omega$ around $\pi/2$ or $3\pi/2$.  While
{\it Spitzer} is now entering the end of its cryogenic lifetime, future
facilities might be able to achieve similar precision. Moreover, as
new suitable systems are discovered with very hot atmospheres
\citep[pM class planets,][]{Fortney2007b} observations of secondary
transits might be feasible from the ground
\citep{Lopez-Morales2007a}.

Finally, we note that since $f_t=\pm\pi/2$ when $\omega=0$ or $\pi$,
measuring the change in $\Delta t$ complements measurements of
changes in primary transit duration $D$, in the sense that when one
effect is not operating  the other one generally is.

\section{Assessing the Different Contributions to $\dot{\omega}$}
\label{sec:distinguish}

If a change in $\omega$ is detected using any of the methods described
above it would be interesting to determine its likely cause.  As
mentioned in \S~\ref{sec:prec}, the contribution to $\dot{\omega}$
from a second planet or tidal deformations can be of comparable
magnitude to that caused by GR, making the origin of a potential
detection of a change in $\omega$ unclear unless the presence of a
perturber and the magnitude of tidal effects can be assessed. 
In this section we address three points. First, we discuss how may one
assess if part of a detected change in $\omega$ comes from a
perturber. Secondly, we show that the a terrestrial mass ($ M_\oplus
\lesssim M \lesssim 10$) perturber may be detectable through the
induced secular changes in $\omega$ over the value expected from GR
and tidal deformations by using transit and radial velocity
observations. Finally, we assess the relative contributions of GR and
tidal deformations for the case where perturbers do not contribute
significantly to $\dot{\omega}$.

\subsection{Constraining Potential Contributions to $\dot{\omega}$
from $\dot{\omega}_{\rm perturber}$.}
\label{ssec:constr_pert}

First, we note that using a precise radial velocity curve with $N_{\rm
obs} \sim 100$ we can probe the presence of companions with $M
\gtrsim 15 M_\oplus$ \citep{Narayan2005a}. As shown in \S~\ref{sec:prec}, 
lower mass companions can still contribute significantly to
$\dot{\omega}$, so it would be useful to have further information
about possible companions in order to better interpret a measured
$\dot{\omega}$.
The presence of a second exoplanet in the system will not only cause
secular changes in $\omega$ as discussed in \S~\ref{sec:prec}
(equation~\ref{eq:wdot_pert}), but it may cause significant
transit-time {\it variations} even for perturbers with masses
comparable to the earth \citep{Agol2005a,Holman2005a}.  The
transit-time variations will be on the order of seconds to minutes,
depending on the orbital parameters and mass of the perturber.
Constant monitoring of transiting exoplanets such as will be performed
by {\it Kepler} will allow to identify systems that show significant
transit time variations.

We note that the presence of a perturber will not only change $\omega$
but it may also affect other orbital elements such as the orbit
inclination
\citep{Miralda-Escude2002a}. These small changes in inclination have indeed
been proposed as a method of detecting terrestrial mass
companions in near-grazing transit systems which are especially
sensitive to them \citep{Ribas2008a}. Following the formalism
presented in \S6 of \citet{MurrayDermott} we expect that secular
changes to $e$ by a perturber satisfy $|d e / dt| \lesssim (3/16) n
(a/a_2)^3 (M_2/\mstar) e_2$, where $a_2, e_2$ and $M_2$ are the
orbital semi-major axies, eccentricity and mass of the perturber
respectively. For $e_2=0.1$, $a_2 = 2a$ and $M_2=3\times10^{-6}\mstar$
we have that $de/dt \lesssim 4 \times 10^{-6}$ yr$^{-1}$ for a system
with $P=4$ days. This is too small for changes in $e$ to be detected
with radial velocities in scales of tens of years and so we conclude
that secular changes in eccentricities induced by perturbers will in
general not be useful to infer the presence of perturbers.

Summarizing, the best way to try to constrain contributions of
$\dot{\omega}_{\rm perturber}$ to a detected change in $\omega$ is to
consider transiting systems posessing extensive photometric
monitoring, allowing to probe the existence of transit time
variations. For near-grazing systems, the same photometric monitoring
would additionally allow to probe for small changes in $i$ due to a
perturber.

\subsection{Using Secular Variations in $\omega$ to Detect
Terrestrial Mass Planets.}
\label{ssec:tpl}

A very interesting possibility raised by the secular variation in
$\omega$ induced by a ``perturber'' is to use these variations in
order to infer the presence of terrestrial mass planets. This
possibility was studied by \citet{Miralda-Escude2002a} and then
followed-up by \citet{Heyl2007a}. Secular variations in $\omega$,
especially through their effect on transit durations
(\S~\ref{subsec:D}), can offer an interesting complement to transit
time variations as a means of detecting terrestrial mass planets in
the upcoming {\it Kepler} mission.

As we have seen in \S~\ref{sec:prec}, the precession due to GR and
tidal deformations can be of comparable magnitude to that induced by a
terrestrial mass perturber. It is germane to ask then to what extent
will the uncertainty in the expected value of \dotw and
$\dot{\omega}_{\rm tide}$ limit the detectability of a perturber. 

We start by considering \dotw. The fractional uncertainty in \dotw is
given by $(\sigma(
\dotw) / \dotw)^2 = (\sigma( \mstar) / \mstar)^2 + (\sigma( a) / a)^2 +
(\sigma( P) / P)^2 + 4e^2 (\sigma( e) / (1-e^2))^2$, where we have
ignored correlations. The fractional uncertainty in $P$ is generally
negligible, while the other quantities can be typically of the order
of a few percent. We will therefore assume that $\sigma(\dotw) / \dotw
\sim 10\%$. It follows that to detect an excess precession caused by a
perturber we need at least that $\dot{\omega}_{\rm perturber} - \dotw
\gtrsim 0.3\dotw$.  It is easy to see that if $\dot{\omega}_{\rm
perturber} \gtrsim \dotw$ and $\dot{\omega}_{\rm perturber}$ is itself
detectable, i.e.  $\dot{\omega}_{\rm perturber} >
3\sigma_{\dot{\omega}}$, where $\sigma_{\dot{\omega}}$ is the
uncertainty in the measured $\dot{\omega}$, then the uncertainty in
the expected $\dotw$ will not spoil the significance of the detection.
Using the equations presented in \S~\ref{sec:prec} we get that
\be
     \frac{\dot{\omega}_{\rm perturber}}{\dotw} = \frac{3.8}{(1-e^2)}
     \left(\frac{M_\odot}{\mstar}\right)^2
     \left(\frac{a}{a_2}\right)^3
     \left(\frac{a}{0.05\mbox{AU}}\right)
     \left(\frac{M_2}{M_\oplus}\right).
\ee
It is clear from this expression that for typical values of $\mstar
\sim M_\odot$, $a \sim 0.05$ AU, $e \lesssim 0.5$ there will be values
of $a_2$ for which we have $\dot{\omega}_{\rm perturber} > \dotw$ and
for which the uncertainty in the expected precession from GR will not
be a limiting factor in detecting terrestrial mass perturbers.

We consider now $\dot{\omega}_{\rm tide}$. The fractional uncertainty
considering all parameters excepting $k_{2,p}$ and ${\cal T}$ and ignoring
correlations is $(\sigma(\dot{\omega}_{\rm tide}) / \dot{\omega}_{\rm
tide})^2 = (\sigma( P) / P)^2 + (5\sigma( \rpl/a) / (\rpl/a))^2 +
(\sigma(
\mpl/\mstar) / (\mpl/\mstar))^2 + (f^\prime(e)
\sigma(e)/f(e))^2$. We have expressed this in terms of the ratios
$\mpl/\mstar$ and $\rpl/a$ as these quantities are more robustly
determined observationally. As was the case above, the uncertainties
in the variables can be typically a few percent and we therefore
assume that $\sigma(\dot{\omega}_{\rm tide}) /
\dot{\omega}_{\rm tide} \sim 10\%$. 
Using the equations presented in \S~\ref{sec:prec} we get that
\be
     \frac{\dot{\omega}_{\rm perturber}}{\dot{\omega}_{\rm tide}} = 
     \frac{1.85}{{\cal T} f(e)k_{2,p}}
     \left(\frac{M_\odot}{\mstar}\right)^2
     \left(\frac{a}{a_2}\right)^3 
     \left(\frac{a}{0.05\mbox{AU}}\right)^5
     \left(\frac{R_J}{\rpl}\right)^5
     \left(\frac{\mpl}{M_J}\right)  
     \left(\frac{M_2}{M_\oplus}\right).
\ee
In the case of $\dot{\omega}_{\rm tide}$ we cannot measure the apsidal
motion constant $k_2$. As discussed in \S~\ref{sec:prec}, the value of
$k_2$ for a giant planet is expected to be close to the extreme value
of a uniform sphere, so we can conservatively assume $k_{2,p}=0.75$, a
value that maximizes the expected $\dot{\omega}_{\rm tide}$. For a
close-in Jupiter we have ${\cal T} \approx 1$. Using these values and
following the same reasoning as for \dotw, it is clear from the
expression above that for typical values of $\mstar
\sim M_\odot$, $a \sim 0.05$ AU, $e \lesssim 0.5$, $\rpl \sim R_J$,
$\mpl \sim M_J$, there will be values of $a_2$ for which we have
$\dot{\omega}_{\rm perturber} >\dot{\omega}_{\rm tide} $ and for which
the uncertainty in the expected precession arising from tidal
deformations will not be a limiting factor in detecting terrestrial
mass perturbers. Note that due to the strong dependence on $a$, tides
may become a limiting factor for very close-in systems.

We consider now the detectability of $\dot{\omega}_{\rm perturber}$
with {\it Kepler} using the change in the transit duration $D$. As
discussed in \S~\ref{subsec:D}, a 1-minute sampling of a Jupiter-mass
system with {\it Kepler} for a $P=5$ days planet orbiting a $V=12$
solar-type star will achieve a precision of $\sigma_D \sim 1.5$ sec
for a 0.15 days transit duration. We therefore set a change of 6 sec
over 4 years to constitute a detectable $\delta D$ for this system,
which translates into a detectable $\dot{\omega}$ of
$\dot{\omega}_{\rm detect} = 6 / (D
\Delta t (\dot{\omega}^{-1}d\ln D/dt))$, where $\Delta t = 4$ years and
we assume a central transit (see Equation~\ref{eq:dlnD}).

\begin{figure}
\epsscale{0.6}
\plotone{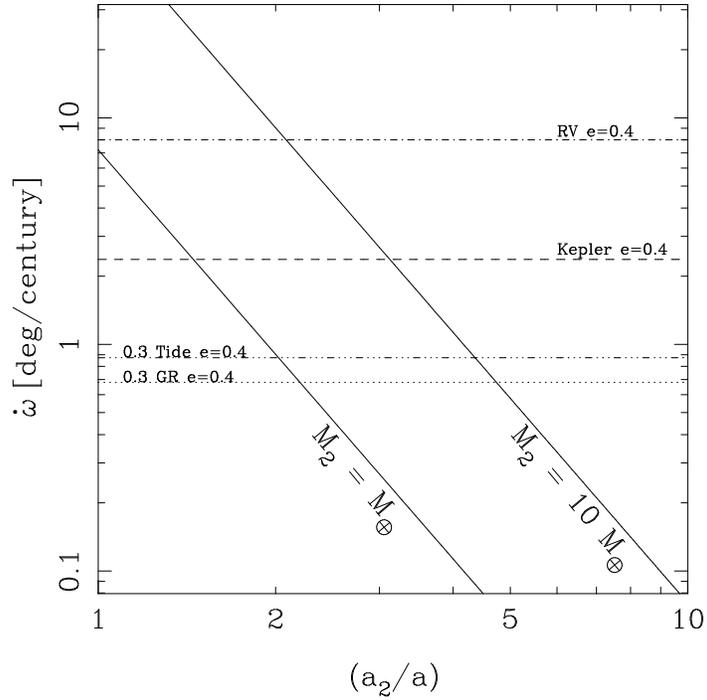}
\caption{
	The solid lines show the expected value of $\dot{\omega}_{\rm
        perturber}$ as a function of $a_2/a$ for $M_2=M_\oplus$ and
        $M_2=10M_\oplus$, assuming $a=0.05$ AU, $\mstar=M_\odot$ and
        $f_t=0.5\pi$. The dashed line marks the lower value of
        $\dot{\omega}$ that is detectable with {\it Kepler} for
        $e=0.4$ via the changes in transit duration. The dotted line
        marks the value of $0.3\dotw$ for the same eccentricity, while
        the dash-dot-dot line marks $0.3 \dot{\omega}_{\rm
        tide}$. Finally, the dot-dashed line marks the lower values of
        $\dot{\omega}$ that would be detectable with 100 radial
        velocities at each of two epochs separated by 5 years, for a
        perturber to a super-massive $M_1=10M_J$ primary planet
        orbiting an inactive solar-mass star (see text for more
        details).  \label{fig:plan_kep} }
\end{figure}

We show in Figure~\ref{fig:plan_kep} the expected value of
$\dot{\omega}_{\rm perturber}$ as a function of $a_2/a$ assuming
$a=0.05$ AU, $\mstar=M_\odot$, and $f_t=0.5\pi$ as solid lines, one
for $M_2=M_\oplus$ and another for $M_2=10M_\oplus$. The dashed line
marks the values of $\dot{\omega}_{\rm detect}$ for $e=0.4$.  The
dotted lines mark the values of $0.3\dotw$ for the same eccentricity,
while the dash-dot-dot line marks $0.3\dot{\omega}_{\rm tide}$ This
figure shows that over a range of values of $(a_2/a)$ {\it Kepler} may
be able to detect the presence of additional super-Earths ($M_\oplus
\lesssim M \lesssim 10M_\oplus$) if the primary planet has a favorable
impact parameter and true anomaly at the time of transit.

Finally, we consider the detectability of $\dot{\omega}_{\rm
perturber}$ using radial velocities alone. We assume the primary
planet is a super-massive system of $\mpl = 10 M_J$, with the rest of
the star and orbital parameters as assumed above. The results of
\S~\ref{subsec:rv} show that with  100 radial velocity observations at
each epoch a precession of $\sim 8$ deg/century could be detected (at
$3\sigma$) in 5 years if the system has $e=0.4$ and orbits an inactive
star. We mark the detectable level in this case with radial velocities
as dot-dashed line in Figure~\ref{fig:plan_kep}. From this Figure we
can see that if enough observations are in hand, perturbing planets
with masses $\sim 10 M_\oplus$ may be detectable with radial
velocities alone for the case of super-massive systems orbiting
inactive stars.

We conclude that measuring the precession of $\omega$ caused by a
perturber above the value predicted by GR and tidal effects may lead
to detection of additional planets with both transit and radial
velocity observations (for transits, see also
\citet{Miralda-Escude2002a,Heyl2007a}). The examples given above are
illustrative only; a detailed analysis is beyond the scope of this
work. Of special interest will be to estimate the yield of
terrestrial-mass planets expected from {\it Kepler} by measuring
secular changes in transit durations.

\subsection{The Relative Magnitude of \dotw and $\dot{\omega}_{\rm tide}$.}
\label{ssec:rel_mag}

If there are no perturbers causing significant changes in $\omega$,
these changes will be caused by a combination of GR and tidal
deformations. We now briefly consider the relative magnitudes of these
effects.  Using the equations presented in \S~\ref{sec:prec} we find
that
\be
     \frac{\dotw}{\dot{\omega}_{\rm tide}} = 
     \frac{4.8}{{\cal T} (1-e^2) f(e)}
     \left(\frac{0.1}{k_{2,p}}\right)     
     \left(\frac{a}{0.05\mbox{AU}}\right)^4
     \left(\frac{R_J}{\rpl}\right)^5
     \left(\frac{\mpl}{M_J}\right).     
\ee
For expected values of $k_{2,p} \sim 0.25$ and assuming $\rpl \sim
R_J$, $\mpl \sim M_J$, $a\sim 0.05$ and $e\lesssim 0.5$ we see that
the values of \dotw and $\dot{\omega}_{\rm tide}$ are comparable. Note
that there is a very strong dependence on $a$. For smaller values of
$a$, a regime where the precession due to GR would offer better
chances of being observable for eccentric systems, the contribution
from tidal deformations may quickly become the dominant source of
precession. The fact that the tidal contributions to the precession
are expected to be generally significant limits the ability to
directly extract the precession due to GR.
The magnitude of the precession caused by tidal deformations is
uncertain due to the need to know the planetary structure.  Given that
the expected precession due to GR is unambigous, in systems where the
precession due to tidal deformations is expected to be dominant {\it
and} detectable, we might be able to learn about the tidally driven
precession of $\omega$ by subtracting the effects of GR from a
measured $\dot{\omega}$.

\section{Analysis of Specific Super-Massive Systems}

As illustration of the material discussed above, we analyze now in
some detail two currently known systems that due to their large
measured velocity amplitudes have the potential to have a measurable
$\dot{\omega}$ using all the techniques presented above: HAT-P-2 b
\citep{Bakos2007a,Loeillet2008a} and XO-3 b
\citep{Johns-Krull2008a}.

\subsection{HAT-P-2 b}

HAT-P-2 b is an especially interesting system due to its very high
eccentricity, high mass of $\approx 9 M_{J}$ and close orbit
($a=0.0677$ AU) to its host star. As shown in Table~\ref{tab:prec} its
value of \dotw is $\sim 2^\circ$/century, while the expected value of
$\dot{\omega}_{\rm tide}$ is $\sim 0.25$ deg/century assuming
$k_{2,p}=0.25$.  Unfortunately it has a high level of stellar jitter,
with estimates ranging from 17 m sec$^{-1}$ to 60 m sec$^{-1}$
\citep{Bakos2007a,Loeillet2008a}. So even though this system has
$K\sim$ 1000 m sec$^{-1}$ it would require an unrealistically large
number of observations ($\sim 10^4$) per epoch in order to make its
expected \dotw detectable with radial velocities.

The value of $\omega$ derived for HAT-P-2 b is $1.05\pi$ and we
therefore would not expect to see significant variations in the time
between primary and secondary transits in case the secondary transit
was observed (see Figure~\ref{fig:dDt}).
With $\omega = 1.05\pi$ the true anomaly at the time of transit will
be $f_t \approx 0.45 \pi$, or $1.55\pi$. Given that the impact
parameter is $b\approx 0$ we get from Equation~\ref{eq:dlnD} that
$|d\ln D/dt| \sim 1.67\times10^{-2}$ century$^{-1}$.  Using the fact
that $D= 0.15$ days we expect then a change of $\sim 21$ sec in $D$
over 10 years due to GR.  This difference could be readily detected
with high precision photometric observations that determine $D$ to
within a few seconds such as is possible with {\it HST}.  The
observations currently available constrain $D$ only to within $\sim 3$
mins \citep{Bakos2007a}, and so no first epoch suitable to measuring
changes in $D$ is yet in hand.
 
\subsection{XO-3 b}
\label{ssec:XO3}

As shown in Table~\ref{tab:prec} XO-3 b has the largest predicted
\dotw of all currently known exoplanets with $e>0.1$. While its
eccentricity is not as high as that of HAT-P-2 b, it is more massive
and orbits closer to its host star XO-3 (also known as
GSC~03727-01064). Asuming $k_{2,p}=0.25$, the expected value of
$\dot{\omega}_{\rm tide}$ is $\sim 11$ deg/century, about three times
as much as the contribution from GR.

In order to assess the detectability of $\dot{\omega}$ with radial
velocities for XO-3 b we need to know $\sigma_{\rm jitter}$, but
unfortunately the precision of the radial velocity measurements
presented in \citet{Johns-Krull2008a} is too coarse to allow a
determination of this quantity ($\sigma_{\rm obs} \gtrsim 100$ m
sec$^{-1}$). Based on the spectral type F5V and $v\sin i=18.5\pm 0.2$
km sec$^{-1}$ of XO-3 b \citep{Johns-Krull2008a} we can expect it to
have a rather high value of stellar jitter $\sigma_{\rm jitter}
\gtrsim 30$ m sec$^{-1}$ \citep{Saar1998a}. Therefore, and just as is
the case for HAT-P-2 b, we do not expect $\dot{\omega}$ to be
detectable with radial velocity observations due to the expected
jitter.
  
The value of $\omega$ derived for XO-3 b is consistent with 0 and we
therefore would not expect to see variations in the time between
primary and secondary transits in case the secondary transit was
observed (see Figure~\ref{fig:dDt}). As $\omega \approx 0$, the true
anomaly at the time of transit will be $f_t \approx \pi/2$ or
$3\pi/2$. Given that the impact parameter is $b\approx 0.8$ we get
from Equation~\ref{eq:dlnD} that $|d\ln D/dt|
\sim 3.56\times10^{-2}$ century$^{-1}$.  Using the fact that
$D = 0.14$ days we expect then a change of $\sim 43$ sec in $D$ over
10 years. Just as is the case for HAT-P-2 b, this could be
detectable with determinations of $D$ to within a few seconds and
there is no suitable first epoch yet in hand.

\section{Conclusions}

In this work we have studied the observability of the precession of
periastra caused by general relativity in exoplanets. We additionally
consider the precession caused by tidal deformations and planetary
perturbers, which can produce a precession of comparable or greater magnitude.
We consider radial velocities and transit light curve observations and
conclude that for some methods precessions of the magnitude expected
from GR will be detectable in timescales of $\sim 10$ years or less
for some close-in, eccentric systems.  In more detail, we find that:

\begin{enumerate}

\item For transiting systems, precession of periastra of the magnitude
expected from GR will manifest itself through detectable changes in
the duration of primary transit (\S\ref{subsec:D}) or through the
change in the time between primary and secondary transits
(\S\ref{subsec:Deltat}) in timescales of $\lesssim 10$ years.  The two
methods are most effective at different values of the true anomaly.  A
determination of the primary transit duration time to $\sim$ a few
seconds and that of the secondary to $\sim$ a minute will lead to
measurable effects. The effects of GR and tidal deformations might
need to be included in the analysis of {\it Kepler} data for
eccentric, close-in systems. The transit duration of near-grazing
systems will be particulary sensitive to changes in $\omega$.

\item Radial velocity observations alone would be able to detect
changes in the longitude of periastron of the magnitude expected from
GR effects only for eccentric super-massive ($K \sim 1000$ m
sec$^{-1}$) exoplanets orbiting close to a host star with a low-level
of stellar jitter (\S\ref{subsec:rv}). For the detection to be
statistically significant, on the order of $100$ precise radial
velocity observations are needed at each of two epochs separated by
$\sim 20$ years.

\item Measurements of the change over time of the period between 
primary transits is not currently a method that will lead to a
detection of changes in $\omega$ of the magnitude expected from GR
(\S\ref{subsec:dPdt}). Previous works have shown that measuring the
small difference between the radial velocity period and that of
transits are not sensitive enough to lead to detectable changes due to
GR.

\end{enumerate}

In order to contrast any detected change in the transit duration
(\S\ref{subsec:D}) or the time between primary and secondary
(\S\ref{subsec:Deltat}) to the predictions of a given mechanism one
needs to know the eccentricity and longitude of periastron of the
systems, for which radial velocities are needed (although not
necessarily of the precision required to directly detect changes in
$\omega$ with them\footnote{The eccentricity can be constrained using
transit information alone, see \citet{Ford2008a}}).
Conversely, photometric monitoring of primary transits are useful in
order to elucidate the nature of a detected change in $\omega$ by
probing for the presence of transit time variations. The presence of
the latter would imply that at least part of any observed changes in
$\omega$ could have been produced by additional planetary companions
(\S\ref{sec:distinguish}).

Precession of periastra caused by planetary perturbers and the effects
of tidal deformations can be of comparable magnitude to that caused by
GR (\S\ref{ssec:add_prec}). The effects of tidal deformations on the
precession of periastra in particular may be of the same magnitude or
dominate the total $\dot{\omega}$ in the regime where the GR effects
are detectable (\S\ref{ssec:rel_mag}). While this limits the ability
to directly extract the precession due to GR given the uncertainty in
the expected precession from tides, it might allow to study the
tidally induced precession by considering the residual precession
after subtracting the effects of GR. The latter possibility is
particularly attractive in systems where the tidally induced
precession may dominate the signal.
We note that even without considering the confusing effects of tidal
contributions to the precession, a measurement of \dotw as described
in this work would not be competitive in terms of precision with
binary pulsar studies \citep[see, e.g.,][for a review ]{Will2006a} and
would therefore not offer new tests of GR.

The upcoming {\it Kepler} mission expects to find a large number of
massive planets transiting close to their host stars
\citep{Borucki2003a}, some of which will certainly have significant
eccentricities. Furthermore, systems observed by {\it Kepler} will be
extensively monitored for variations in their transiting time periods
in order to search for terrestrial-mass planets using transit-time
variations. We have shown that modeling of the transit time durations
and further characterization of close-in, eccentric systems might need
to take into account the effects of GR and tidal deformations as they
will become detectable on timescales comparable to the 4-year lifetime
of the mission, and certainly on follow-up studies after the mission
ends. We have also shown that planetary companions with super-Earth
masses may be detectable by {\it Kepler} by the change in transit
durations they induce (\S~\ref{ssec:tpl}). Additionally, well sampled
radial velocity curves spanning $\gtrsim 5$ years may also be able to
detect companions with super-earth massses by measuring a change in
$\omega$ over the expected GR value for the case of super-massive,
close-in systems orbiting inactive stars (\S~\ref{ssec:tpl}).

\acknowledgements

We would like to thank the anonymous referee for helpful suggestions
and Dan Fabrycky and Andr\'as P\'al for useful
discussions. G.B. acknowledges support provided by the National
Science Foundation through grant AST-0702843.

\bibliographystyle{apj}
\bibliography{allrefs}

\end{document}